# Dephasing by a Continuous-Time Random Walk Process

Daniel M. Packwood and Yoshitaka Tanimura

**Abstract**

Stochastic treatments of magnetic resonance spectroscopy and optical spectroscopy require evaluations of functions like $\left\langle \exp\left(i\int_0^t Q_s ds\right) \right\rangle$, where $t$ is time, $Q_s$ is the value of a stochastic process at time $s$, and the angular brackets denote ensemble averaging. This paper gives an exact evaluation of these functions for the case where $Q$ is a continuous-time random walk process. The continuous time random walk describes an environment that undergoes slow, step-like changes in time. It also has a well-defined Gaussian limit, and so allows for non-Gaussian and Gaussian stochastic dynamics to be studied within a single framework. We apply the results to extract qubit-lattice interaction parameters from dephasing data of P-doped Si semiconductors (data collected elsewhere), and to calculate the two-dimensional spectrum of a three level harmonic oscillator undergoing random frequency modulations.

## 1. Introduction

Relaxation of a ensemble to equilibrium is the central problem of non-equilibrium statistical mechanics. One particular example of relaxation is dephasing, which is of considerable interest to the quantum information community [1,2]. As an example of dephasing, consider an ensemble of qubits that have each been put into a coherent state. Because each qubit experiences different random interactions with its environment, the phase of the qubits will drift apart with time and the ensemble will no longer exhibit quantum mechanical properties [3]. Coherent states also arise in spectroscopy, for example, between the time at which a molecule is excited by an external field and when it emits a photon [4]. Random modulations of the phases of these molecules by interactions with their environments means that photons end up being emitted at a variety of frequencies. This causes the net emission of the ensemble at a particular frequency to decay with time, and this decay partly determines the shape of the spectral lines that are measured in the experiment [5,6,7]. This paper is about a stochastic *continuous time random walk* model of dephasing. We will present the key equations and show how to evaluate them. We will then apply the model to study qubit dephasing in solid-state environments and to calculate two-dimensional spectra of a harmonic oscillator. This paper follows a previous paper, in which the properties of the continuous-time random walk were studied in detail and the relaxation equations evaluated approximately in certain limits [8].

Stochastic models of dephasing were pioneered by Kubo and Anderson in the 1960s [5,9]. In these models, the microscopic physics of the system-environment interaction are modelled by adding a stochastic process to the relevant frequencies of the system. While stochastic models lack the atomistic detail of other models, they are of great practical use because they can be applied without too much effort and have intuitive physical interpretations. If we bring an ensemble of systems out of equilibrium by means of an external field, then to first order the average phase of the ensemble relaxes to equilibrium according to the function

$$F(t) = \left\langle \exp\left( i \int_0^t Q(r) dr \right) \right\rangle, \qquad (1)$$

where $Q(t)$ is the value of a stochastic process at time $t$ and $\langle \ \rangle$ denotes ensemble averaging over the possible realisations of $Q(t)$ [5,6,7]. $F(t)$ goes by a variety of names in the literature, including the characteristic functional [10,11], the decoherence function [12] the Kubo-Anderson correlation function [13] and, depending on the context, the first-order response function [7]. We will call $F(t)$ the phase relaxation function. For problems in non-linear spectroscopy, in which special experimental schemes are used, we need to compute higher order relaxation functions such as

$$F(t) = \left\langle \exp\left( ic_1 \int_{t_0}^{t_1} Q(r) dr + ic_2 \int_{t_2}^{t_3} Q(r) dr + \cdots + ic_{n-1} \int_{t_{n-1}}^{t_n} Q(r) dr \right) \right\rangle, \qquad (2)$$

where $t_0 \leq \cdots \leq t_n$ and $c_1,\ldots,c_{n-1}$ are real constants [4]. Thus, solving stochastic models of phase relaxation comes down to evaluating a phase relaxation function. The phase relaxation function in (1) has been evaluated for the case of a stationary Gaussian process (an Ornstein-Uhlenbeck process) [5], and the cumulant expansion technique allows for evaluation of higher-order relaxation functions such as (2). Gaussian processes apply when the interactions between the system and environment occur very frequently on the experimental time scale. For interactions that take place on a slower time scale, we have stationary Markov jump processes (also known as the random telegraph process). The phase relaxation function in equation (1) (but apparently not the one in (2)) has been evaluated and applied for a variety of such cases [5,9, 11,12,13,14,15,16].

Like the stationary Markov jump processes, the continuous time random walk process (CTRW) describes relatively infrequent system-environment interactions. It runs as follows. Choose an initial value $Q_0$ for the process and hold it for a random duration $K_1$. Then, select a random number $X_1$ and hold the process at the value $Q_0 + X_1$ for a random duration $K_2$. Then, select another random number $X_2$, and so on. The key difference between the CTRW and the Markov jump processes is that it evolves through a series of small steps $X_1, X_2, \ldots$, rather than by making unrestrained leaps across the state space. This might be appropriate for e.g., the effective magnetic field felt by a qubit on a lattice of nuclear spins. In this situation, two of the spins exchange polarisation at random points in time, and this causes an small change in the effective magnetic field felt by the qubit [17,18,19]. The CTRW is also non-stationary, and so describes an environment that is held out of equilibrium by an external field [20]. An attractive feature of the CTRW is that the limit $K_1, K_2, \ldots \to 0$ yields a well-defined Gaussian process (the Wiener process), and so we can investigate the transition from non-Gaussian CTRW dynamics to Gaussian dynamics within a single model [8,21,22,23]. Physically, this limit corresponds to changes in the system-environment interaction occurring more and more frequently on the time-scale of the experiment.

This paper is organised as follows. Second 2 reviews the CTRW and shows how the relaxation functions in equations (1) and (2) can be evaluated. Section 3 applies the relaxation function to qubit dephasing in solid state environments, and Section 4 computes the two-dimensional spectra of a simple harmonic oscillator system. Conclusions and general comments are left for Section 5.

## 2. Relaxation Function for a CTRW

We denote random variables by capital, italic Roman letters, and use subscript *t* to indicate the value of a stochastic process at time *t*. Expected values (ensemble averages) are indicated by $E(\ )$ or $\langle \ \rangle$. This section begins with a short review of the continuous-time random walk. The phase relaxation function is derived in part ii.

*i. The continuous-time random walk*

A stochastic process can be visualised in terms of an imaginary particle making a motion along the real numbers. Let the position of the particle at time 0 be $Q_0$. The CTRW model supposes that the particle stays at this position up to a random time $U_1$, where it takes a step of size $X_1$ and moves to a new position $Q_0 + X_1$. The particle stays in this position up to another random time $U_2$, where it undergoes a change of size $X_2$ and moves to position $Q_0 + X_1 + X_2$. And so on. If the waiting times $K_1 = U_1, K_2 = U_2 - U_1, \ldots$ are exponential random variables, i.e., they all have the distribution

$$P(K_i < k) = 1 - e^{-\lambda k}, \tag{3}$$

where $1/\lambda = E(K_1) = E(K_2) = \cdots$ is the average length of the waiting times, then the position of the particle at time *t* is

$$Q_t = Q_0 + \sum_{i=1}^{N_t} X_i. \tag{4}$$

In (4), $N_t$ is the value of a Poisson process at time *t*. (4) is only really useful when the random variables $X_1, X_2, \ldots$ are independent and have the same distribution (i.e., they are *iid* random variables). By (3), the CTRW is a Markov process [24].

It will be helpful to derive the characteristic function of the CTRW at time *t*. We therefore wish to compute

$$\phi_{Q_t}(v) = E(e^{ivQ_t}), \tag{5}$$

where *v* is a real constant. We can do this *via* conditional expectation, namely

$$E(e^{ivQ_t}) = E(E(e^{ivQ_t} | N_t)). \tag{6}$$

(6) says that the characteristic function can be computed by first supposing that $N_t$ is a non-random constant, computing the expectation of $\exp(ivQ_t)$, and then averaging the result over all $N_t$. The first step gives

$$E\left(\exp(ivQ_0)\exp\left(iv\sum_{j=1}^{N_t} X_j\right) \bigg| N_t\right) = \phi_{Q_0}(v)\phi_X^{N_t}(v), \tag{7}$$

where $\phi_{Q_0}(v)$ and $\phi_X(v)$ are the characteristic functions of the initial position and steps $X_1, X_2, \ldots$, respectively. (7) results from the fact that $X_1, X_2, \ldots$ are independent and have the same distribution. We have also assumed that the initial condition and $X_1, X_2, \ldots$ are independent. Averaging (7) over $N_t$ then gives

$$E\left(e^{ivQ_t}\right) = \phi_{Q_0}(v) \sum_{n=0}^{\infty} \phi_X^n(v) P(N_t = n)$$
$$= \phi_{Q_0}(v) \exp\left(-\lambda t \left(1 - \phi_X(v)\right)\right) \tag{8}$$

The first line in (8) is from the formula $E(f(N_t)) = \sum_{n=0}^{\infty} f(n) P(N_t = n)$ [25]. The second was obtained by substituting in the Poisson distribution, $P(N_t = n) = \exp(-\lambda t) \lambda^n t^n / n!$. (8) will be useful once the distribution of the initial coordinate and of the step sizes are specified.

For the applications in this paper we will work with the *continuous-jump CTRW*. Here, the distribution of the step sizes $X_1, X_2, \ldots$ is uniform on $(-M, M)$, where $M > 0$. The characteristic function of the step sizes is then [26]

$$\phi_X(v) = \frac{\sin(vM)}{vM}. \tag{9}$$

For a continuous-jump CTRW, it can be shown that as $\lambda \to \infty$ and successive jumps of the particle occur more and more frequently, the distribution of $Q_t - Q_0$ converges to the distribution of $W_t$, a Wiener process (a non-stationary Gaussian process) at time *t*, providing that $M$ takes on the value

$$M = \sqrt{3/\lambda}, \tag{10}$$

somewhere close to the limit [8]. (10) is a renormalisation of the constant *M*. Thus, a particle undergoing a continuous jump CTRW with very frequent, very small jumps, can be regarded as one undergoing a Wiener process. A physical interpretation of the renormaliser in (10) was given in reference [8].

*ii. The phase relaxation functions*

We will consider three variants of the phase relaxation function. Type 1 is a minor generalisation of (1), namely

$$F_1(t) = E\left(\exp\left(ic\int_0^t Q_r dr\right)\right), \qquad (11)$$

where $c$ is a real constant. Type 1 appears in first-order perturbative treatments of relaxation, such as in linear response theory [5,6,7]. For the continuous-jump CTRW, this works out to be (see the appendix)

$$F_1(t) = \phi_{Q_0}(ct) e^{-\lambda t} \exp\left(\frac{\lambda}{cM} \operatorname{Si}(t;cM)\right), \qquad (12)$$

where $\phi_{Q_0}$ is the characteristic function of the initial condition and $\operatorname{Si}(t;cM)$ is a special case of the sine integral,

$$\operatorname{Si}(t;cM) = \int_0^t \frac{\sin(cMr)}{r} dr. \qquad (13)$$

The sine integral cannot be computed analytically, however it is easy to evaluate numerically. It is here computed with the trapezium rule, using a partition of the interval $(0,t)$ of size 10,000. Convergence is achieved with partitions of size around 7000. Figure 1 plots the Type 1 relaxation function $F(t)$ for various $\lambda$, $M = 1$, and with non-random initial conditions (i.e., $Q_0 = 0$, $\phi_{Q_0}(ct) = 1$). The trends are the same as those predicted by approximate formulae in the previous paper [8]. Namely, that relaxation becomes faster as $\lambda$ and $M$ increase. For very small $\lambda/M$, the second exponential in (12) is close to 1 and the decay is exponential. For larger values of $\lambda/M$, the decay of the curves is slightly irregular. This irregular behaviour has been seen in similar studies of the two-state Markov jump process by Kitajima and co-workers [11]. If the initial condition $Q_0$ is a Gaussian random variable rather than a non-random constant, then

$$\phi_{Q_0}(v) = \exp(-v^2\sigma^2/2), \tag{14}$$

where $\sigma$ is the standard deviation of $Q_0$. We will use this initial condition throughout the rest of the paper. $\sigma = 0$ corresponds to a non-random initial condition. This random initial condition enhances the exponential decay by adding a quadratic term $-\sigma^2 t^2/2$. This causes the irregular features in Figures 1 to become less apparent (result not plotted).

The Type 2 phase relaxation function is

$$F_2(t_1,t_2) = E\left(\exp\left(ic\int_{t_1}^{t_2} Q_r dr\right)\right), \tag{15}$$

where $t_2 \geq t_1$. This type of phase relaxation function appears in certain non-linear optical problems such as pump-probe spectroscopy [4]. In the appendix we show that

$$F_2(t_1,t_2) = \phi_{Q_0}(c(t_2-t_1))e^{-\lambda t_2}$$
$$\times \exp\left(\lambda\left[\frac{t_1}{t_2-t_1}\frac{\sin(c(t_2-t_1)M)}{cM} + (t_2-t_1)\operatorname{Si}(t_2-t_1;cM)\right]\right). \tag{16}$$

The Type 3 phase relaxation function is particularly relevant to two dimensional spectroscopy:

$$F_3(t_1,t_2,\ldots,t_m) = E\left(\exp\left(ic_1\int_0^{t_1} Q_r dr + ic_2\int_{t_1}^{t_2} Q_r dr + \cdots + ic_2\int_{t_{m-1}}^{t_m} Q_r dr\right)\right), \tag{17}$$

where $t_m \geq t_{m-1} \geq \cdots \geq t_1 \geq 0$ and $c_1, c_2, \ldots$ are constants that can be set to zero. There does not seem to be an obvious way to generalise the derivation of the Type 1 and 2 phase relaxation function to the Type 3 relaxation function in (17). However, we can derive a recursive formula that can be computed to any desired degree of accuracy without too much effort, namely

$$F_2(t_1, \ldots, t_n) = \lim_{n \to \infty} \phi_{Q_0}(t_m B_1) \prod_{k=1}^{n} \phi_{\Delta Q}(t_m B_k) \tag{18}$$

where

$$\phi_{\Delta Q}(t_k B_k) = \exp\left(-\frac{\lambda t}{n}\left(1 - \phi_X(t_k B_k)\right)\right) \tag{19}$$

with $\phi_X$ given by (9), and

$$B_k = \frac{1}{n} \sum_{i=k}^{n} \beta(t_i) \tag{20}$$

and $\beta(r)$ is such that

$$t_{k-1} \leq r < t_k \Rightarrow \beta(r) = c_k. \tag{21}$$

For a given $n$, (18) can be evaluated by first computing $\phi_{Q_0}(t_m B_1) \phi_{\Delta Q}(t_m B_1)$, then multiplying the result by $\phi_{\Delta Q}(t_m B_2)$, and so on, up to $\phi_{\Delta Q}(t_m B_n)$. Convergence is similar to numerical integration of the sine integral, and is achieved at around $n = 8000$. We will use $n = 10000$ in the following calculations. For a given $t_m$, this calculation takes a few seconds on a modern desktop computer.

**3. Application to qubit dephasing**

Consider an ensemble of non-interacting spins in a magnetic field $B_0$ parallel to the z-axis. The spins each have eigenstates $|0\rangle$ and $|1\rangle$, and we use $\omega$ to denote the frequency of the transition $|0\rangle \rightarrow |1\rangle$. $\omega$ is proportional to **$B_0$**. The magnetic field causes the spins vectors to precess about the z axis. A magnetic field **$B_1$** circularly polarised in the xy plane (from a radiofrequency field) with frequency $\omega$ is then applied to the spins, causing the spin vectors to shift into the xy plane. The spin vectors are then rotating around the xy plane with angular frequency $\omega$. Phase decoherence occurs as the angular frequencies of the ensemble are modulated due to interactions with the surrounding environment, and eventually the spin vector fall out of alignment with each other. We can map the rotation in the xy plane my mapping the motion onto the unit circle by means of the complex number $\exp\left(i\int_0^t \omega(s)ds\right)$. Letting $\omega(t) = \omega + Q_t$ and taking expectations gives the linear response function

$$R(t) = e^{i\omega t} F_1(t), \qquad (22)$$

where $F_1(t)$ is a Type 1 relaxation function with $c = 1$ (see (11)). It turns out that the free induction decay measured in a magnetic resonance experiment is proportional to the real part of $R(t)$.

The case where $Q$ is a continuous-time random walk can be used to model dephasing for qubits in a crystalline material such as silicon. In a crystal under high magnetic fields, dephasing is primarily caused by random exchange of spins across the crystalline lattice. For example, consider phosphorous-doped silicon. In this material, the unpaired electrons of the phosphorous atoms can be regarded as qubits. These electrons are relatively localised on the P atom (i.e., their wave functions do not extend far beyond the P atom and into the Si lattice [27]). Natural crystals of silicon-28 contain 4.7 % [29]Si, which has a nuclear spin of 1/2. Two nearby [29]Si atoms can exchange their polarisation, and in turn this polarisation can travel around the lattice. When this polarisation comes close enough to any qubits, hyperfine interactions between the [29]Si and electron will cause transition frequency $\omega$ of the qubit to fluctuate [17]. We will model this as follows. Consider a single qubit (i.e., a single electron from a phosphorous atom). Starting from time 0, we suppose that the silicon lattice remains unchanged up to time $U_1$, where two adjacent [29]Si nuclei in the lattice exchange polarisation and hence cause a small change $X_1$ in the resonance frequency of the qubit. The silicon lattice then remains unchanged up to time $U_2$, where a random change in the lattice causes a change of size $X_2$ in the resonance frequency of the qubit. And so on. We can model the stochastic part of the resonance frequency of the qubit as a continuous time random walk if we assume exponential waiting times between successive polarisation exchange events. The continuous time random walk model is very approximate compared to other theoretical treatments of polarisation exchange (for example, [18,19,27,28,29]). It seems reasonable under the following circumstances. *i*. The $^{29}$Si spins are widely spaced apart. *ii*. Polarisation exchange between one pair of $^{29}$Si is independent of polarisation exchange between any other pair of $^{29}$Si. *iii*. Each adjacent $^{29}$Si pair in the lattice has an equal chance to flipping at each of the times $U_1, U_2, \ldots$. Assumption *i* means that the change in the resonance frequency of the qubit due to a flip of one $^{29}$Si is not cancelled by the opposite change due to the flip of the other spin. This other spin is too far from the qubit to have a significant effect. The concentration of $^{29}$Si in silicon is very small, so this assumption is reasonable. Assumption *ii* means that the random variables $X_1, X_2, \ldots$ are independent. Assumption *iii* means that $X_1, X_2, \ldots$ are uniform random variables between $-M$ and $+M$, where $M$ is a positive constant. $X_1, X_2, \ldots$ will be close to $\pm M$ if the spin that flips is close to the qubit, and will be closer to zero for spins that are further away. $M$ will be related to the nuclear spin-electron spin hyperfine splitting constant, although we will not investigate the connection here.

Figure 2a compares a plot of the Type 1 relaxation function for a continuous-jump random walk with experimental data of photon echo decay of a natural Si crystal doped with phosphorous (data reported by Tyryshkin *et al.* [17]). We assumed that the bandwidth of the pulse used in the EPR experiment is so narrow that errors in $Q_0$ can be disregarded (i.e., $\sigma = 0$). Note that the oscillating component of the data has been subtracted. From this, we adjusted *M* and *λ* to 15 ms$^{-1}$ and 8 ms$^{-1}$, respectively, to produce the fit in the figure. The agreement is reasonably good, although the relaxation function overestimates the echo decay signal at small times and decays more quickly. The oscillations at small times in the data are an experimental artefact and can be ignored [17]. According to this value of *λ*, the qubit experiences an average of only 8 ms$^{-1}$ x 0.5 ms = 4 changes in the surrounding nuclear spin environment during the course of the experiment. While the hyperfine coupling constant between the unpaired electron on phosphorous and the $^{29}$Si nuclei in crystalline silicon does not appear to have been reported in the literature, the value of *M* appears to be very small. For example, Assali *et al* estimate that the coupling constant between an electron and $^{29}$Si nucleus in a silicon quantum dot to be about 10,000 times larger [30]. However, in this case we would expect the coupling constant to be larger because the electron is not localised on a phosphorous atom and the shape of the quantum dot would cause the $^{29}$Si nuclei to be quite close to the electron. In the case of crystalline silicon, the number of $^{29}$Si nuclei that are in the immediate vicinity of the phosphorous atoms in the lattice might be very small (if most $^{29}$Si were far away from the phosphorous atom, then the coupling would be weakened and *M* would be reduced). This would worth investigating further in experimental studies and computational studies, and theoretical studies might like to investigate the connection between *M* and the P electron-$^{29}$Si nucleus hyperfine coupling constant in more detail.

We can also derive a simple graphical method of estimating *λ* and *M* from experimental dephasing data. Taking the logarithm of the Type 1 relaxation function gives

$$\ln F(t) = -\lambda t + \frac{\lambda}{M} \int_0^t \frac{\sin Mr}{r} dr \ . \tag{23}$$

Using the fact that $\lim_{t \to \infty} \int_0^t \sin(Mr)/r \, dr = \pi/2$, we can approximate (23) for large times:

$$\ln F(t) \approx \frac{\pi \lambda}{2M} - \lambda t \ . \tag{24}$$

Thus, for large $t$ a plot of $\ln F$ against $t$ should yield a straight line with slope $\lambda$ and intercept $\pi\lambda/(2M)$. Figure 2b plots the the logarithm of the data of Tyryshkin *et al* [17], and fits a linear regression line to the data collected beyond 0.4 ms (fitting performed in *R 2.13.1* [31]). The fitting procedure gives $\lambda$ = 11.31 ms$^{-1}$ and *M* = 9.83 ms$^{-1}$. These values are slightly different from those estimated by directly fitting the relaxation function in Figure 2a, but are in the same order-of-magnitude. Equation (24) looks appropriate for making 'ballpark' estimates of the dephasing parameters $\lambda$ and *M* from experimental dephasing data.

As well as the other approximations mentioned earlier, the continuous time random walk approach is restricted to `large' magnetic fields only. For small magnetic fields random exchange of spins across the lattice is supplemented with various other nuclear spin-nuclear spin interactions [18,32]. This also prevents the continuous time random walk model from exploring interesting features of spin dephasing that occur at low magnetic fields, such as periodic revivals of the echo signal in a lattice of nuclei with different Lamor frequencies [18,19,32] and dependences of the echo signal on the orientation of the crystal in the magnetic field [17]. Nonetheless, these shortcomings are more than redeemed by the ease with which the continuous time random walk can be applied to experimental data. Highly accurate and physically detailed models can almost never be applied such a straightforward fashion.

**3. Application to two-dimensional spectroscopy**

The goal of two dimensional spectroscopy is to measure the third-order response function,

$$R_3(t_3, t_2, t_1) = \frac{i}{\hbar^3} \left\langle \mu(t_3) \left[ \mu(t_2), \left[ \mu(t_1), \left[ \mu(0), \rho(-\infty) \right] \right] \right] \right\rangle, \tag{25}$$

where $\mu(t_k)$ is the electric dipole operator of the system of interest at time $t_k$ and $\rho(-\infty)$ the density operator of the system at the initial time $-\infty$ [4,33]. The electric dipole operator is in the interaction picture, and the times $t_1$, $t_2$ and $t_3$ indicate different interaction times with an external electromagnetic field. In a two-dimensional spectroscopy experiment, the field interactions represented by $\mu(0), \mu(t_1), \mu(t_2)$ and $\mu(t_3)$ correspond to different laser pulses at times $0, t_1, t_2$ and $t_3$ striking an ensemble of systems. Between times 0 and $t_1$ the ensemble is relaxing from an excited state, and we can Fourier transform the macroscopic polarisation during this time to obtain a (one dimensional) spectrum. Similarly, between times $t_2$ and $t_3$ the ensemble is relaxing again and we can obtain from that another (one dimensional) spectrum. For a fixed $t_2 - t_1$, the two-dimensional spectrum is the two-dimensional Fourier transform of the relaxation dynamics over the intervals $(0, t_1)$ and from $(t_2, t_3)$. See [33] for more details.

Consider a quantum harmonic oscillator with three levels. In increasing order of energy, the levels are $|0\rangle$, $|1\rangle$ and $|2\rangle$. The $|1\rangle \leftarrow |0\rangle$ and $|2\rangle \leftarrow |1\rangle$ transition energies are both $\hbar\omega$. We will suppose that $\omega(t) = \omega_0 + Q_t$, where $\omega_0$ is a constant and $Q_t$ is the value of a continuous-time random walk at time $t$. If we expand the commutators in (25) then we obtain a variety of terms. Each term can be represented by a double-sided Feynman diagram from Figure 3. The complex conjugates of the diagrams are not shown. By integrating the von Neumann equation ($d\rho/dt = -(i/\hbar)[H_0, \rho]$, where $\rho$ is the density matrix of the system and $H_0$ is the unperturbed Hamiltonian of the system) we can find expressions for the evolution of the density matrix elements between the laser pulses. These work out to be

$$\begin{pmatrix} \dot\rho_{00} & \dot\rho_{01} & \dot\rho_{02} \\ \dot\rho_{10} & \dot\rho_{11} & \dot\rho_{12} \\ \dot\rho_{20} & \dot\rho_{21} & \dot\rho_{22} \end{pmatrix} = -i \begin{pmatrix} 0 & -\omega(t)\rho_{01} & -2\omega(t)\rho_{02} \\ \omega(t)\rho_{10} & 0 & -\omega(t)\rho_{12} \\ 2\omega(t)\rho_{20} & \omega(t)\rho_{21} & 0 \end{pmatrix}.$$

(26)

Using the shorthand

$$Z_{\pm\pm\pm}^{a_1 a_2 a_3} = \left\langle \exp\left( \pm i a_1 \int_0^{t_1} Q_r dr \pm i a_2 \int_{t_1}^{t_2} Q_r dr \pm i a_3 \int_{t_2}^{t_3} Q_r dr \right) \right\rangle$$

(27)

and integrating (26), the terms corresponding to the diagrams in Figure 3 work out to be

$$R_1^L = \exp(-i\omega t_1 - i\omega(t_3 - t_2))Z_{-0-}^{101}$$
$$R_1^R = \exp(-i\omega t_1 + i\omega(t_3 - t_2))Z_{-0+}^{101}$$
$$R_2^L = \exp(i\omega t_1 - i\omega(t_3 - t_2))Z_{+0-}^{101}$$
$$R_2^R = \exp(i\omega t_1 + i\omega(t_3 - t_2))Z_{+0+}^{101}$$
$$R_3^L = \exp(i\omega t_1 + 2i\omega(t_2 - t_1) + i\omega(t_3 - t_2))Z_{+++}^{121}$$
$$R_3^R = \exp(i\omega t_1 - i\omega(t_3 - t_2))Z_{+0-}^{101}$$
$$R_4^L = \exp(-i\omega t_1 - 2i\omega(t_2 - t_1) - i\omega(t_3 - t_2))Z_{---}^{121}$$
$$R_4^R = \exp(-i\omega t_1 - i\omega(t_3 - t_2))Z_{-0-}^{101}$$

(28)

where we have set all dipole transition moments to one. See [33] for further details. The third-order response function is

$$R_3(t_3, t_2, t_1; t_2 - t_1) = \frac{i}{\hbar^3}\left(\sum_{j=1}^{4}(R_j^L + R_j^R) - \left(\sum_{j=1}^{4}(R_j^L + R_j^R)\right)*\right)$$

(29)

The argument $t_2 - t_1$ after the semicolon indicates that in a two-dimensional spectroscopy experiment $t_2 - t_1$ is treated as a parameter. The factors $Z_{\pm\pm\pm}^{a_1 a_2 a_3}$ are Type 3 phase relaxation functions and can be computed with (18).

Figure 4 presents two-dimensional spectra computed from (29). Because the raw spectra are difficult to visualise and involve complex and imaginary parts, we here present them as power spectra. They were calculated with the *fft* subroutine of *R 2.13.0* [31]. We set $t_2 - t_1 = 1$, $M = 1$, $\sigma = 0.1$ and $\omega = 1$ (arbitrary units). The spectra were normalised by dividing through by the height of the peak in the $\lambda = 0.01$ case. The spectra have a cross shape which is typical of many two dimensional spectra. The key observation is that the peak height decreases as $\lambda$ increases and the continuous time random walk approaches the Wiener process in the Gaussian limit. The Wiener process is a non-stationary Gaussian process. For the case of a stationary Gaussian process the two-dimension spectrum of a three-level quantum oscillator is known to vanish completely [34]. The spectra in Figure 6 therefore show a clear difference between non-Gaussian, continuous-time random walk stochastic modulation and Gaussian stochastic modulation. This difference might be useful for experimentalists wishing to spot signatures of non-Gaussian frequency modulation in the laboratory.

## 5. Final Remarks

The contribution of this paper is the evaluation of the phase relaxation function for a continuous time random walk. We evaluated the phase relaxation function analytically in the Type 1 and 2 cases (up to the sine integral), and provided a straightforward numerical algorithm for computing the more general Type 3 case. The results are in equations (12), (16) and (18). This algorithm is so fast and easy to implement that for practical purposes it can be regarded as a 'complete' evaluation of the Type 3 relaxation function. We demonstrated how these relaxation functions can be applied in practice by finding a way to extract qubit-lattice interaction parameters from dephasing data from P-doped silicon semiconductors, and by showing that a strong signal in the two-dimensional spectrum of an oscillator implies strongly non-Gaussian, continuous time random walk-type stochastic frequency modulation.

Prior to this research, there were only two types of stochastic processes for which the relaxation function had been evaluated. Namely, for Gaussian processes [5,9] and for the Markov jump process [10,11,12,13,14,15,16]. Our results add the continuous-time random walk to this list. We evaluated its moments in full in a previous paper, as well as its probability density function (see [8]; note that the characteristic function in equation in (8) of this paper is more useful than the expression for the density derived in the previous paper). These properties, as well as the relaxation function, are probably sufficient for the continuous-time random walk to be applied to most typical problems in non-equilibrium statistical mechanics. One particular advantage of the continuous-time random walk is that it converges to a well-defined Gaussian process (a Wiener process), which allows for non-Gaussian and Gaussian dynamics to be studied within a single model. A possible setback is that the continuous-time random walk and limiting Wiener process are non-stationary, whereas physical applications often require a stochastic process to be stationary. Nonetheless, the continuous-time random walk can still be used in an approximate manner (stochastic processes are only heuristic descriptions an environment anyways), and the non-stationary property may be useful for describing a system in a non-equilibrium environment [20]. Future applications of the continuous time random walk might include more rigorous modelling of qubit dephasing and of the non-linear spectra of more interesting, non-Gaussian systems like the O-H bond of water [35,36,37].

## 6. Appendix: Proof of the phase relaxation function formulas

*Type 1* (equation (12)). If we partition the time interval $(0,t)$ into intervals $(t_0,t_1),(t_1,t_2),\ldots,(t_{n-1},t_n)$, where $t_0 = 0, t_n = t$, and each segment has length $t/n$, then we can approximate the integral in (11) as its Riemann sum:

$$\int_0^t Q_s ds \approx \sum_{k=1}^n Q_{t_k} \frac{t}{n}. \tag{30}$$

The formula is exact in the limit $n \to \infty$. Letting $\Delta Q_{t_k} = Q_{t_k} - Q_{t_{k-1}}$, the sum in (30) can be re-written as

$$\sum_{k=1}^{n} Q_{t_k} \frac{t}{n} = \frac{t}{n} \sum_{k=1}^{n} \left( Q_{t_0} + \Delta Q_{t_1} + \cdots + \Delta Q_{t_k} \right)$$
$$= \frac{t}{n} \left( nQ_{t_0} + n\Delta Q_{t_1} + (n-1)\Delta Q_{t_2} + \cdots + (n-(n-1))\Delta Q_{t_n} \right) \quad (31)$$

Introducing the sequence

$$A = \{a_k\}$$
$$= \{1 - (k-1)/n\}_{k=1}^{n} \quad (32)$$

we can re-write (31) as

$$\int_{0}^{t} Q_s \, ds \approx tQ_{t_0} + t\sum_{k=1}^{n} a_k \Delta Q_{t_k}. \quad (33)$$

Let us note the limit behaviour of the sequence $A$. Explicitly, $A = \{1, 1-1/n, 1-2/n, \ldots, 1/n\}$. Taking the limit gives

$$\lim_{n \to \infty} A = (0,1] \cap \mathbb{Q}, \quad (34)$$

where $\mathbb{Q}$ is the rational numbers. We will use this result soon. Substituting (33) into (11) gives

$$F_1(t) = E\left[ e^{ictQ_{t_0}} \prod_{k=1}^{n} \exp\left( icta_k \Delta Q_{t_k} \right) \right]. \quad (35)$$

Because the time intervals $(t_0,t_1),(t_1,t_2),\ldots$ do not overlap, the sums $\Delta Q_{t_1} = \sum_{i=1}^{N_{t_1}} X_i, \Delta Q_{t_2} = \sum_{i=N_{t_1}+1}^{N_{t_2}} X_i,\ldots$ do not contain any common elements. The increments $\Delta Q_{t_1}, \Delta Q_{t_2},\ldots$ are therefore independent random variables. We will also suppose these increments are independent of the initial frequency $Q_{t_0}$. Moreover, because $X_1, X_2,\ldots$ have the same distribution, and that the time intervals $(t_0,t_1),(t_1,t_2),\ldots$ have the same length, (35) factorises into

$$F_1(t) = \phi_{Q_{t_0}}(ct) \prod_{k=1}^{n} \phi_{\Delta Q_{t_k}}(cta_k), \qquad (36)$$

where $\phi_{Q_{t_0}}(ct) = E(\exp(ictQ_{t_0}))$ and $\phi_{\Delta Q_{t_k}}(cta_k) = E(\exp(icta_k \Delta Q_{t_i}))$ are the characteristic functions of the random variables $Q_{t_0}$ and $\Delta Q_{t_k}$, respectively. $\phi_{\Delta Q_{t_k}}(cta_k)$ can be computed *via* conditional expectation, namely

$$\phi_{\Delta Q_{t_k}}(cta_k) = E\left[ E\left( \exp\left( icta_k \sum_{j=N_{t_{k-1}}+1}^{N_{t_k}} X_j \right) \middle| N_{t_k} - N_{t_{k-1}} \right) \right]. \qquad (37)$$

This works out to be

$$\phi_{\Delta Q_{t_k}}(cta_k) = \exp\left( -\frac{\lambda t}{n}(1-\phi_X(cta_k)) \right). \qquad (38)$$

Substituting (38) into (36),

$$F_1(t) = \phi_{Q_{t_0}}(ct) e^{-\lambda t} \exp\left( \lambda t \left( \frac{1}{n} \sum_{k=1}^{n} \phi_X(cta_k) \right) \right). \qquad (39)$$

Now, according to (34) and the definition of the Riemann integral,

$$\lim_{n\to\infty}\sum_{k=1}^{n}\phi_X(cta_k)(1/n)=\int_0^1\phi_X(ctz)dz. \tag{40}$$

Substituting (40) into (39) gives

$$F_1(t)=\phi_{Q_0}(ct)e^{-\lambda t}\exp\left(\lambda t\int_0^1\phi_X(ctz)dz\right). \tag{41}$$

For the case of the continuous jump CTRW, we can go further and eliminate the variable *z*. Substituting in (9), we obtain

$$F_1(t)=\phi_{Q_0}(ct)e^{-\lambda t}\exp\left(\frac{\lambda}{cM}\int_0^1\frac{\sin(ctMz)}{z}dz\right). \tag{42}$$

Now, let

$$G(t)=\exp\left(\frac{\lambda}{cM}\int_0^1\frac{\sin(ctMz)}{z}dz\right) \tag{43}$$

and differentiate it with respect to *t*:

$$\frac{dG(t)}{dt}=\frac{\lambda}{ctM}\sin(ctM)G(t). \tag{44}$$

Solving (44) and noting that $G(0)=1$ from (43) gives

$$G(t) = \exp\left(\frac{\lambda}{cM}\text{Si}(t;cM)\right) \tag{45}$$

where

$$\text{Si}(t;cM) = \int_0^t \frac{\sin(cMr)}{r}dr. \tag{46}$$

Substituting (45) back into (42) yields (12).

*Type 2* (equation (16)). To avoid confusion between notation, we will briefly re-write the Type 2 relaxation function in (16) as

$$F_2(s_1, s_2) = E\left(\exp\left(ic\int_{s_1}^{s_2} Q_r dr\right)\right) \tag{47}$$

Following the steps for the Type 1 relaxation function, we partition $(s_1, s_2)$ into $n$ intervals of length $(s_2 - s_1)/n$. The corresponding Riemann sum is

$$\int_{s_1}^{s_2} Q_r dr = \sum_{k=1}^{n} Q_{t_k}\left(\frac{s_2 - s_1}{n}\right), \tag{48}$$

where this time $t_0 = s_1 < t_1 < t_2 < \cdots < t_n = s_2$. As in (31) and (33), this can be re-written as

$$\int_{s_1}^{s_2} Q_r dr \approx (s_2 - s_1)Q_{t_0} + (s_2 - s_1)\sum_{k=1}^{n} a_k \Delta Q_{t_k}. \tag{49}$$

Carrying this expression through the steps in the previous proof gives

$$F_2(s_1, s_2) = \phi_{Q_{s_1}}\left(c(s_2 - s_1)\right) e^{-\lambda(s_2 - s_1)} \exp\left(\lambda(s_2 - s_1)\int_0^1 \phi_X\left(c(s_2 - s_1)z\right) dz\right), \quad (50)$$

or, upon substituting in (9) and switching back to the original notation,

$$F_2(t_1, t_2) = \phi_{Q_0}\left(c(t_2 - t_1)\right) e^{-\lambda t_2} \\ \times \exp\left(\lambda\left[t_1 \phi_X\left(c(t_2 - t_1)\right) + (t_2 - t_1)\int_0^1 \phi_X\left(c(t_2 - t_1)z\right) dz\right]\right). \quad (51)$$

For the continuous-jump random walk, this works out to be

$$F_2(t_1, t_2) = \phi_{Q_0}\left(c(t_2 - t_1)\right) e^{-\lambda t_2} \\ \times \exp\left(\lambda\left[\frac{t_1}{t_2 - t_1} \frac{\sin\left(c(t_2 - t_1)M\right)}{cM} + (t_2 - t_1)\operatorname{Si}(t_2 - t_1; cM)\right]\right) \quad (52)$$

*Type 3* (equation (18)). Re-write the integral in equation (17) as

$$ic_1 \int_0^{t_1} Q_r dr + ic_2 \int_{t_1}^{t_2} Q_r dr + \cdots + ic_2 \int_{t_{m-1}}^{t_m} Q_r dr = i \int_0^{t_m} \beta(r) Q_r dr, \quad (53)$$

Where $\beta(r)$ is such that

$$t_{k-1} \leq r < t_k \Rightarrow \beta(r) = c_k \quad (54)$$

By partitioning the interval $(0, t_m)$ into *n* subintervals and approximating equation (53) with its Riemann integral, we can show that (cf. equation (33))

$$\sum_{k=1}^{n} \beta(t_k) Q_{t_k} \frac{t_m}{n} = \frac{t_m}{n} \Big( Q_{t_0} \big( \beta(t_1) + \cdots + \beta(t_m) \big) + \Delta Q_{t_1} \big( \beta(t_1) + \cdots + \beta(t_m) \big)$$
$$+ \Delta Q_{t_2} \big( \beta(t_2) + \cdots + \beta(t_m) \big) + \cdots + \Delta Q_{t_n} \beta(t_m) \Big).$$
(55)

Substituting (55) into (17) and using the *iid* property of the increments gives the result.

## 6. References

[1]   T. Endo, J. Phys. Soc. Jpn **56**, 1684 (1987)

[2]   H. Breuer and F. Petruccione, *The Theory of Open Quantum Systems* (Oxford University Press, New York, 2002)

[3]   M. Stoneham, Physics **2**, 34 (2009)

[4]   S. Mukamel, *Principles of Nonlinear Optical Spectroscopy* (Oxford University Press, New York, 1995)

[5]   R. Kubo, Adv. Chem. Phys. **15**, 101 (1969)

[6]   S. Duttagupta, *Relaxation Phenomena in Condensed Matter Physics* (Academic Press, Orlando, FL, 1987).

[7]   Y. Tanimura, J. Phys. Soc. Jpn. **75**, 82001 (2006)

[8]   D. M. Packwood and Y. Tanimura, Phys. Rev. E. **84**, 61111 (2011)

[9]   P. W. Anderson, J. Phys. Soc. Jpn. **9**, 316 (1954)

[10]  M. Ban, S. Kitajima, K. Maruyama and F. Shibata, Phys. Lett. A. **372**, 351 (2008)

[11]  S. Kitajima, M. Ban and F. Shibata, J. Phys. B: At. Mol. Opt. Phys. **43**, 135504 (2010)

[12]  M. Ban, F. Shibata and S. Kitajima, J. Mod. Opt. **54**, 555 (2007)

[13]  E. Barkai, Y. Jung and R. J. Silbey, Annu. Rev. Phys. Chem. **55**, 457 (2004)

[14]  Y. Jung, E. Barkai and R. J. Silbey, Chem. Phys. **284**, 181 (2002)

[15]  N. A. Sergeev and M. Olszewski, Solid. State. Nucl. Mag. **34**, 167 (2008)

[16]  M. Olszewski and N. A. Sergeev, Z. Naturforsch. **63a**, 688 (2008)

[17]  A. M. Tyryshkin, J. J. L. Morton, S. C. Benjamin, A. Ardavan, G. A. D. Briggs, J. W. Ager and S. A. Lyon, J. Phys.: Condens. Matter **18**, S783 (2006)

[17]   L. Cywinski, W. M. Witzel and S. Das Sarma, Phys. Rev. Lett **102**, 57601 (2009)

[19]   L. Cywinshi, W. M. Witzel and S. Das Sarma, Phys. Rev. B. **79**, 245314 (2009)

[20]   S. Kawai and T. Komatsuzaki, J. Chem. Phys. **134**, 114523 (2011).

[21]   D. M. Packwood, J. Phys. A: Math. Theor. **43**, 465001 (2010).

[22]   D. M. Packwood. Hokkaido University Technical Report Series in Mathematics **151**, 168 (2012)

[23]   R. Metzler and J. Klafter, Phys. Rep. **339**, 1 (2000).

[24]   P. G. Hoel, S. C. Port amd C. J. Stone, *Introduction to Stochastic Processes* (Waveland Press, Prospect Heights, IL, 1972)

[25]   I. Miller and M. Miller, *John E. Freund's Mathematical Statistics with Applications* (Pearson Prentice Hall, Upper Saddle River, NJ, 2004)

[26]   J. Galambos, *Advanced Probability Theory* (Marcel Dekker, New York, NY, 1995).

[27]   W. M. Witzel, R. de Sousa and S. Das Sarma, Phys. Rev. B. **72**, 161306 (2005)

[28]   W. M. Witzel and S. Das Sarma, Phys. Rev. B. **74**, 35322 (2006)

[29]   R. de Sousa, and S. Das Sarma, Phys. Rev. B. **68**, 115322 (2003)

[30]   L. V. C. Assali, H. M. Petrilli, R. B. Capaz, B. Koiller, X. Hu and S. Das Sarma, Phys. Rev. B. **83**, 165301 (2011)

[31]   R Development Core Team, R: A Language and Environment for Statistical Computing (R Foundation for Statistical Computing, Vienna, Austria, 2010), http://www.R-project.org/.

[32]   H. Bluhm, S. Foletti, I. Neder, M. Rudner, D. Mahalu, V. Umansky and A. Yacoby, Nature Phys. **7**, 109 (2011)

[33]   P. Hamm and M. Zani, *Concepts and Methods of 2D Infrared Spectroscopy* (Cambridge University Press, Cambridge, U.K., 2011)

[34]   A. Ishizaki and Y. Tanimura, J. Chem. Phys **125**, 84501 (2006)

[35]   T. Steinel, J. B. Asbury, S. A. Corcelli, C. P. Lawrence, J. L. Skinner, and M. D. Fayer, Chem. Phys. Lett. **386**, 295 (2004).

[36]   T. I. C. Jansen, D. Cringus, and M. S. Pshenichnikov, J. Phys. Chem. A **113**, 6260 (2009).

[37]   S. Roy, M. S. Pshenichnikov, and T. L. C. Jansen, J. Phys. Chem. B **115**, 5431 (2011).

**Figures**

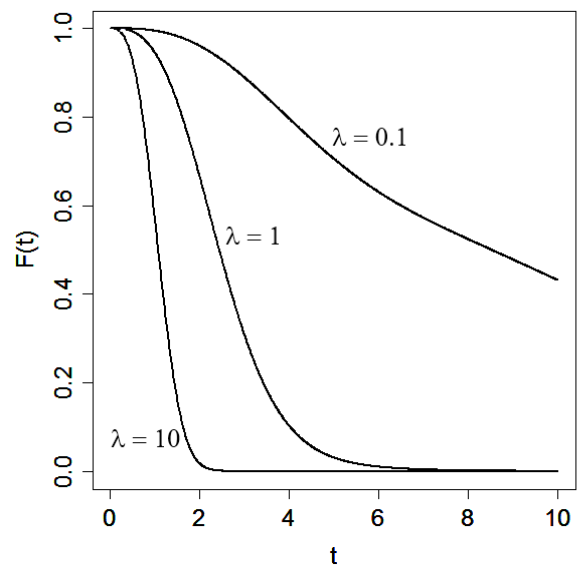

Figure 1. Plot of the phase relaxation function for the continuous-time random walk (12) for $M = 1$ and $\sigma = 0$.

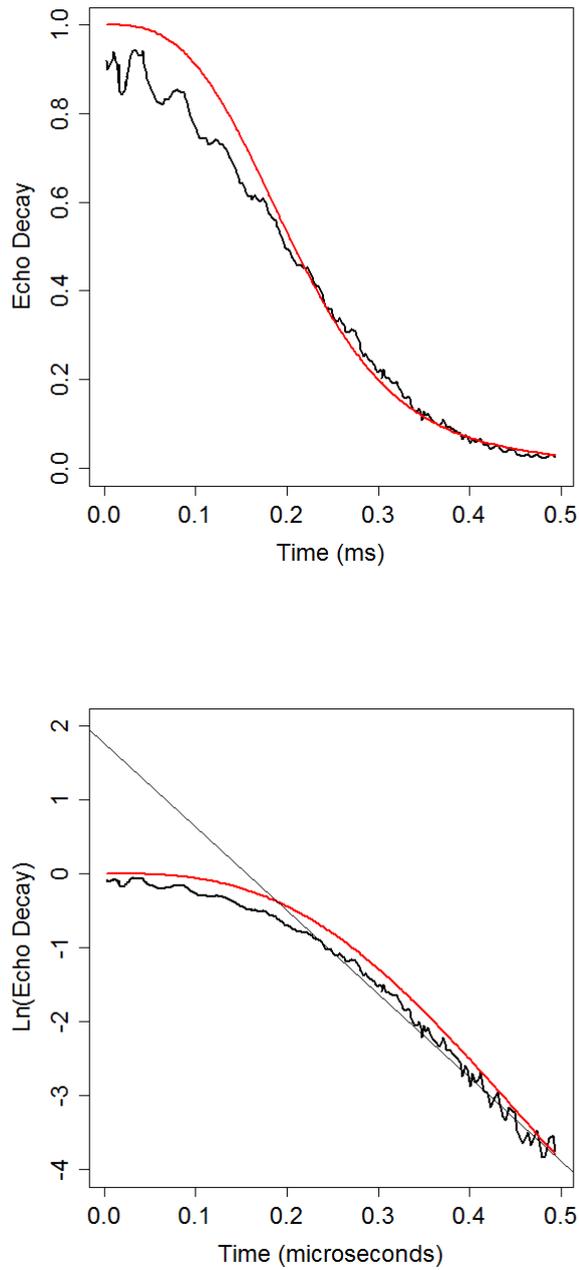

Figure 2. Top: A comparison of the phase relaxation function (12) (red line) with experimental dephasing data from a natural silicon crystal doped with phosphorous, using $\sigma = 0$, $M = 15$ ms$^{-1}$ and $\lambda = 8$ ms$^{-1}$. Bottom: The log of the experimental data from the top figure with (24) (thin black line). (24) was fit with least squares regression to data in the range of 0.4 ms to 0.5 ms, and predicts $M \sim 10$ ms$^{-1}$ and $\lambda \sim 11$ ms$^{-1}$. The red line is a plot of the log of the phase relaxation function (12) using these parameters. Experimental data is from [Tyrshykin].

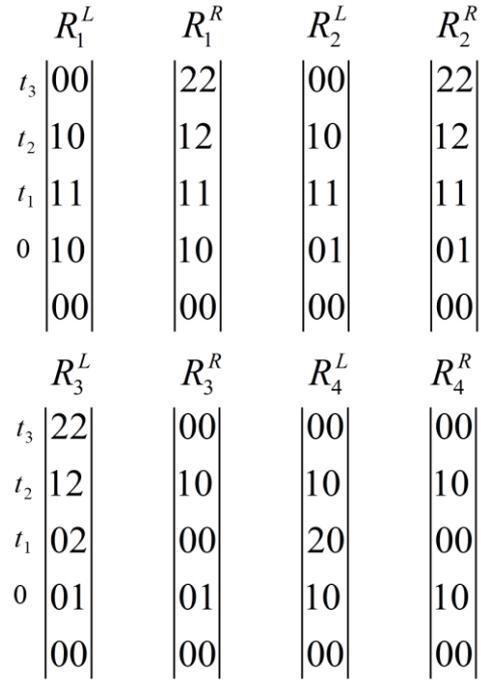

Figure 3. Double-sided Feynman diagrams for a system with eigenstates $|0\rangle$, $|1\rangle$ and $|2\rangle$. Each diagram corresponds to a term from the expansion of the commutators of (25) (complex conjugate terms are not included). Time runs from bottom to top. The two numbers *A* and *B* enclosed by the two vertical lines represent the ket and bra of the density matrix, respectively.

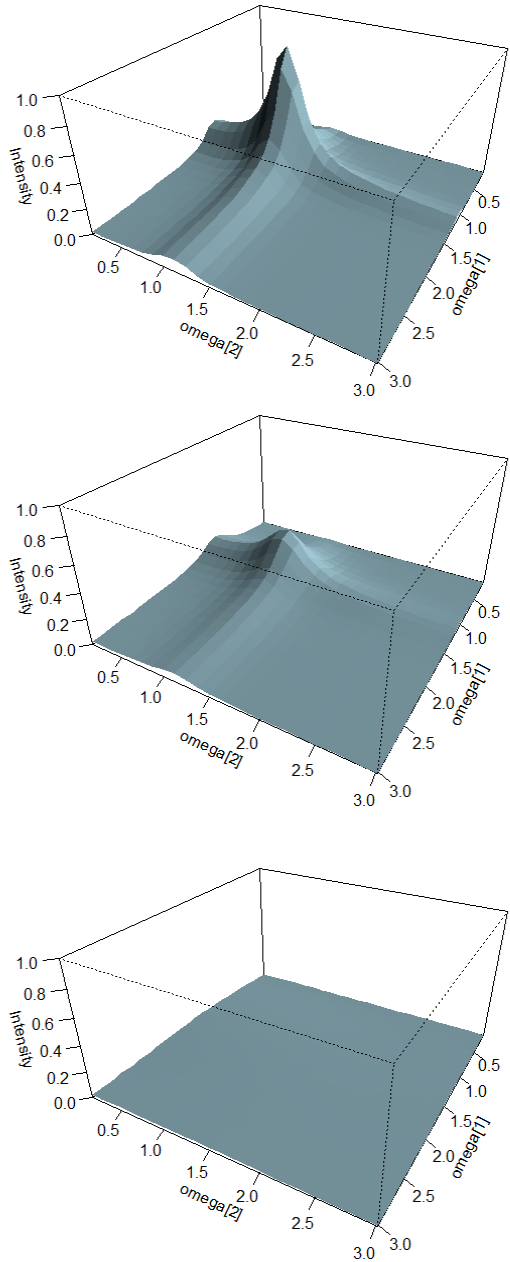

Figure 4. Two dimensional spectra for a three level harmonic oscillator computed from (29), with $t_2 - t_1 = 1$, $M = 1$, $\sigma = 0.1$ and $\omega = 1$ (in arbitrary units) and $\lambda = 0.01$ (top), 0.1 (middle) and 1 (bottom). In the figures, omega[1] and omega[2] are the frequencies from the Fourier transform of (29) over the intervals $t_2 - t_1$ and $t_3 - t_2$, respectively.